\documentclass[a4paper, amsfonts, amssymb, amsmath, reprint, showkeys, nofootinbib,twocolumn,dvipsnames,superscriptaddress]{revtex4-2} 
\usepackage{mypackage}
\usepackage{algorithm}
\usepackage{algpseudocode}
\usepackage{float}
\usepackage{tcolorbox}
\usepackage{enumitem} 
\usepackage{subfigure}
\floatstyle{ruled}
\restylefloat{algorithm}
\begin{document}

\title{Unconditional Authentication in Quantum Key Distribution via Hybrid Entangled Physical Unclonable Functions}

\author{Nicolas Laurent-Puig}
    \email[Correspondence email address:]{nicolas.laurent-puig@lip6.fr}
    \affiliation{Sorbonne Université, CNRS, LIP6, 4 Place Jussieu, Paris F-75005, France}
\author{Mina Doosti}
    \affiliation{School of Informatics, University of Edinburgh, 10 Crichton Street, Edinburgh EH8 9AB, UK}
\author{Adriano Innocenzi}
    \affiliation{Sorbonne Université, CNRS, LIP6, 4 Place Jussieu, Paris F-75005, France}
\author{Eleni Diamanti}
    \affiliation{Sorbonne Université, CNRS, LIP6, 4 Place Jussieu, Paris F-75005, France}

\date{\today} 
\begin{abstract}
Quantum Key Distribution (QKD) enables Information-Theoretically Secure (ITS) key exchange, robust even against future quantum computing threats. However, a fundamental limitation of QKD is the requirement for an authenticated classical channel, which necessitates a pre-shared secret key. In this work, we address this challenge by adopting a Hybrid Entangled Physical Unclonable Function (PUF) protocol for authentication. We demonstrate that this PUF-based method generates an ITS initial key under minimal explicit hardware assumptions. This approach allows us to experimentally perform a fully ITS-authenticated entanglement-based QKD protocol that relies solely on such assumptions, effectively eliminating the need for pre-shared secrets. This represents a significant step towards the practical realization of quantum network protocols using lightweight, readily available hardware assumptions, without weakening security guarantees.
\end{abstract}

\maketitle
\textit{\textbf{Introduction—}}In 1984, Bennett and Brassard established that two parties can distribute a secret key with Information-Theoretic Security (ITS), immune even to adversaries with unbounded computational power~\cite{bennett1984quantum}. This paradigm, namely Quantum Key Distribution (QKD), is based on fundamental laws of quantum mechanics to detect eavesdropping in the quantum channel~\cite{gisin2002quantum, scarani2009security}. However, while the key exchange offers unconditional security, the protocol strictly requires the classical channel to be authenticated. Without authentication, QKD is susceptible to Man-in-the-Middle attacks, rendering the theoretical security of the quantum layer moot~\cite{mosca2018cybersecurity,renner2023debate}. This authentication requirement presents a significant ``chicken-and-egg'' problem: to generate a secret key, the communicating parties must already share a secret key for authenticating the classical channel, typically via a Wegman-Carter authentication scheme~\cite{wegman1981new}. Standard solutions to this bottleneck often compromise the ITS promise of QKD. Although Post-Quantum Cryptography (PQC) algorithms allow for public-key distribution that can be used for authentication, they rely on unproven computational hardness assumptions~\cite{bernstein2017post}. This reintroduces the very same computational vulnerability that QKD aims to eliminate. This has motivated research into alternative assumptions to support new solutions, in particular hardware-based assumptions like Physical Unclonable Functions (PUFs)~\cite{pappu2002physical}. PUFs exploit the stochastic, unclonable physical disorder introduced during device manufacturing to create a unique hardware fingerprint. Although the literature has largely focused on classical PUFs, recent work by Arapinis \textit{et al.}~\cite{arapinis2021quantum} has formally extended this concept to the quantum domain, defining the security properties of Quantum PUFs (qPUFs). However, authentication protocols based on fully quantum PUFs remain technically demanding: they typically require strong quantum capabilities from the device and verifier, making them difficult to realise with near-term technology. This motivates hybrid PUF constructions, which retain the unclonability and hardware-specific behaviour of PUFs while reducing the quantum requirements to a more practical level. Building upon this direction, Goswami \textit{et al.}~\cite{2504.11552} developed a novel hardware architecture, the Hybrid Entangled PUF (HEPUF). The HEPUF leverages the unique hardware fingerprint of the PUF while exploiting quantum properties such as local indistinguishability. By synthesizing these primitives, the HEPUF facilitates a full authentication protocol that preserves the information-theoretic security standards of the quantum communication network without relying on computational assumptions.


In this Letter, we propose and report on the experimental implementation of HEPUF authentication integrated within an Entanglement-Based (EB) QKD protocol. Theoretically, we develop an efficient protocol based on~\cite{2504.11552}. Experimentally, we realize this protocol using a high-fidelity Bell state source operating at telecom wavelength, ensuring compatibility with fiber-optic infrastructures. We demonstrate secret-key distribution over a total channel attenuation reaching 50~dB, maintaining a Quantum Bit Error Rate (QBER) consistently below 0.6\,\%. Furthermore, we assess the protocol under two scenarios: a general scenario demonstrating the versatility of the HEPUF authentication protocol, and a high-efficiency implementation leveraging the intrinsic properties of EB QKD. To the best of our knowledge, this is the first experimental realization of hybrid hardware-based authentication protocols.

\textit{\textbf{Theoretical Protocol—}}We start by presenting the complete authenticated QKD protocol, which integrates two distinct subroutines: the HEPUF authentication subroutine, inspired by the so-called online protocol of~\cite{2504.11552}, and the Quantum Key Distribution subroutine outlined in Protocol~\ref{protocol}. The execution of the overall protocol is iterative; by selecting appropriate initialization parameters, we determine the minimum number of iterations required for the HEPUF authentication to generate a secret key of sufficient length. Once established, this initial key serves as the fundamental resource for the following subroutine. For the more general scenario that we consider in this work, the key is utilized in the EB QKD subroutine both as the seed for the bases choice and for a Wegman-Carter (WC) authentication scheme. This ensures Information-Theoretic Security for the classical post-processing channel. In a scenario that is more efficient in terms of secret-key rate, we declare the initial bases choices publicly (which is possible for the EB QKD protocol) and only use the key for the WC scheme. Although both scenarios offer equivalent security, the first approach provides a versatile framework for implementing a broader class of protocols with specific requirements, such as prepare-and-measure QKD.\\ 
\texttt{Hybrid PUF Authentication.} The subroutine utilizes the HEPUF architecture detailed in End Matter~\ref{sec:HEPUF Architecture}, which operates on the entangled basis set $\{\ket{\Phi^+}, \ket{\Psi^-}\}$. We define two parties: Bob (the Prover) and Alice (the Verifier). The Authentication subroutine proceeds as follows: the Verifier builds a classical Challenge-Response Database from the HEPUF (Mode 0), and transmits it with the HEPUF (Mode 1) to the Prover. The modes correspond to fixed behaviors of the HEPUF and are detailed in End Matter~\ref{sec:HEPUF Architecture}. Upon receiving a challenge $x_i$ from the Verifier, the Prover inputs the challenge into the HEPUF, generating a response $y_i$. This response is partitioned into two bits, $y_i = y_i^1 \parallel y_i^2$. The second bit, $y_i^2$, determines the quantum state preparation: the Prover generates the corresponding Bell state (either $\ket{\Phi^+}$ or $\ket{\Psi^-}$) and transmits one half of the bipartite system to the Verifier over the quantum channel. The state generation and selection occur inside the device, and the HEPUF (Mode 2) only outputs the resulting mixed state. Subsequently, the Prover performs a projective measurement on their local subsystem on a basis dictated by the first bit, $y_i^1$, and discloses the measurement outcome via the classical channel. The Verifier measures their received qubit and cross-references the correlation between the measurement outcomes with the expected values stored in the Database. To maximize efficiency, we implement a batched execution mode: the Verifier issues a set of random challenges, and for each challenge, the Prover generates $m$ states. This block processing authenticates a response 2$m$-bit string of $f(x) = y_i^1 \parallel y_i^2 $ bits per challenge. By repeating this subroutine for $N$ challenges we establish an initial length $2Nm$. To account for inherent physical biases in the PUF, we apply randomness extraction to the raw output. This process yields a uniform, authenticated initial key $K_{\text{HEPUF}}$ of length $L \leq 2Nm$.\\
\texttt{Quantum Key Distribution.}
The QKD subroutine is based on the framework established by Tomamichel and Leverrier~\cite{Tomamichel2017largelyself}. We implement the protocol in two operational scenarios. Alice and Bob share a bit string to determine their measurement bases, which may be kept secret or made public depending on the specific protocol requirements. Based on their initial parameters, the parties compute the total required length for the initial secret key—accounting for classical authentication and, if applicable, the secret basis selection. The combined key is defined as $K_{\text{HEPUF}}$ = $K_{\text{QKD}}\parallel K_{\text{AUT}}$ and is generated via the previously described HEPUF subroutine. Utilizing $K_{\text{AUT}}$, the parties authenticate all subsequent classical communication via a Wegman-Carter (WC) scheme.
They share $n$ copies of a bipartite quantum state $\rho_{AB}$ and perform local measurements in the chosen bases, yielding the raw-key strings $X$ (Alice) and $Y$ (Bob). Subsequently, the parties perform Parameter Estimation to assess the security of the channel. Alice reveals a subset of her outcomes over the classical channel, allowing Bob to compute the Quantum Bit Error Rate (QBER). If the estimated QBER exceeds a predefined threshold $\Delta_{\text{tol}}$, the protocol is aborted. Upon successful Parameter Estimation, the parties proceed to Error Correction (or Information Reconciliation). Alice computes a syndrome $T$ from her remaining raw key and transmits it to Bob. Using $T$ and $Y$, Bob generates an estimator $\hat{X}$ of Alice's key. To verify the correctness of the reconciliation, Alice sends a hash of her raw key $h(X)$; if Bob's computed hash $h(\hat{X})$ does not match, the protocol is aborted. Finally, the parties perform Privacy Amplification to eliminate potential information leakage. Alice and Bob apply a randomness extractor based on a universal$_2$ hash family to derive identical final secret keys $K_A$ and $K_B$ of length $\ell$. We are adopting a strict security model in which all classical communication requires authentication. 

Having established the general framework, we now delineate two distinct scenarios determined by the trust assumptions and the desired protocol universality. The first scenario (Scenario 1) leverages the versatility of the HEPUF scheme. Here, Alice and Bob pre-select their measurement bases using the secure output string of the HEPUF. Since the HEPUF response is information-theoretically secure and exclusive to the honest parties, the measurement bases sequence remains completely concealed from the eavesdropper. By preventing basis-dependent attacks, this approach renders the authentication protocol compatible with a wide range of quantum cryptographic schemes. Notably, the proposed HEPUF construction is compatible with any other QKD architecture, including Prepare-and-Measure, as well as other cryptographic primitives, like Quantum Digital Signatures~\cite{gottesman2001quantum}. The second scenario (Scenario 2) is optimized for key rate efficiency, assuming the implementation of an EB QKD protocol. In this case, the security is guaranteed by non-local correlations rather than the secrecy of the state preparation. Consequently, Alice and Bob can publicly disclose their measurement bases over the classical channel. This relaxes the demand on the HEPUF throughput, as the secret key is consumed solely for authentication, thereby increasing the overall efficiency of the system. 

We can now evaluate the security performance of the protocol to determine the required resource overhead for authentication, characterized by the sample size per round, $m$, and the total number of repetition rounds, $N$. We quantify security by the parameter $\epsilon$ and impose a uniform bound on both the HEPUF Authentication and QKD subroutines, defined as $\epsilon_{\text{HEPUF}} < \epsilon$ and  $\epsilon_{\text{QKD}} < \epsilon$.

In the following analysis, we derive the security bound for the HEPUF authentication, $\epsilon_{\text{HEPUF}}$. In our framework, $\epsilon_{\text{HEPUF}}$ quantifies the maximum probability that an unbounded adversary successfully compromises the authentication key. The overall protocol is secure against Quantum Polynomial Time (QPT) adversaries including forgery adversaries in the pre-protocol stage. The failure probability is bounded by:
\begin{equation}
\epsilon_{\text{HEPUF}} \leq \left(\frac{1}{2} + \delta\sqrt{\frac{1+4\delta^2}{2}}\right)^m,
\end{equation}
where $\delta$ represents the bias of the underlying weak classical PUF~\cite{Marconot2020} and we recall that $m$ denotes the number of states produced per round. This relation demonstrates that the security of the HEPUF depends intrinsically on the device-specific parameter $\delta$ and the sample size $m$.
\begin{tcolorbox}[colback=white,colframe=black,sharp corners, boxrule=0.8pt, title=\small\textbf{Protocol 1: Authenticated QKD}]
\small
\textbf{Goal:} Establish secure key $K_{\text{A}} =K_{\text{B}} $ of length $l$.

\textbf{Inputs:} HEPUF, Quantum \& Public Channels.

\noindent\rule{\textwidth}{0.4pt}

\begin{enumerate}
    \item \textbf{Hybrid PUF \label{protocol}Authentication~\cite{2504.11552}}
    \begin{enumerate}
        \item \textbf{Setup:} Alice employs HEPUF (Mode~0) to create Database $\mathcal{D}=\{(x_i, y_i)\}$ with $y_i=y_i^1 \parallel y_i^2$. She securely sends HEPUF (Mode~1) to Bob.
        
        \item \textbf{Authentication:} 
        \begin{itemize}
            \item Alice sends random challenge $x_i \in \mathcal{D}$ to Bob over a public channel and Bob computes $y_i$ using the HEPUF (Mode~1).
            \item Bob prepares $\ket{\Phi^+}$ if $y_i^2=0$ or $\ket{\Psi^-}$ if $y_i^2=1$.
            \item Bob sends one qubit to Alice and measures his local qubit in $Z$ if $y_i^1=0$ or in $X$ if $y_i^1=1$ using the HEPUF (Mode~2) and announces the outcome.
        \end{itemize}

        \item \textbf{Verification:} Alice measures using basis $y_i^1$ from $\mathcal{D}$. She verifies correlations: outcomes must be identical if $y_i^2=0$ or opposite if $y_i^2=1$.
    \end{enumerate}

    \vspace{2pt} 

    \item \textbf{Quantum Key Distribution (QKD)~\cite{Tomamichel2017largelyself}}
    \begin{enumerate}
        \item \textbf{Key Partition:} Alice sends a seed $S_{\text{INT}}$. Parties compute $K_{\text{HEPUF}}$ = $K_{\text{QKD}}\parallel K_{\text{AUT}}$ from valid $y_i$ strings.
        \item \textbf{Measurement:} Parties share $n$ copies of $\rho_{AB}$, measuring in bases derived from $K_{\text{QKD}}$. They obtain raw strings $X$ (Alice) and $Y$ (Bob).
        \item \textbf{Parameter Estimation:} Alice reveals a subset of $X$. If QBER $> \Delta_{\text{tol}}$ they abort.
        \item \textbf{Error Correction:} Alice sends syndrome $T$. Bob uses $Y$ and $T$ to generate estimator $\hat{X}$.
        \item \textbf{Error Verification:} Alice sends hash $h(X)$. Bob checks if $h(\hat{X}) = h(X)$. If not, they abort.
        \item \textbf{Privacy Amplification:} Two-universal hashing yields final shared key $K_{\text{A}} = K_{\text{B}}$.
    \end{enumerate}
\end{enumerate}
\textbf{Channel Authentication:} Parties authenticate the classical channel using $K_{\text{AUT}}$ (Wegman-Carter).
\end{tcolorbox} From the Leftover Hash Lemma~\cite{renner2006securityquantumkeydistribution}: 
\begin{equation}
    l \leq H_{min}^{\epsilon_{\text{stat}}}(V\mid E) - 2\log_2\left(\frac{1}{\epsilon_{\text{stat}}}\right), 
\end{equation}
where $\epsilon_{\text{stat}}$ accounts for the smoothing parameter (statistical error) in the min-entropy estimation, $V$ is a variable and $E$ is the information held by the adversary.
We can deduce the length of the extractable initial secret key, $l$:
\begin{equation}
    l \leq 2Nm \left[ -\log_2\left(\frac{1+2\delta}{2}\right) \right] - 2\log_2\left(\frac{1}{\epsilon_{\text{stat}}}\right).
\end{equation}

Finally, the total security parameter for the QKD subroutine, $\epsilon_{\text{QKD}}$, is derived following the finite-size security proofs established in~\cite{scarani2009security,Tomamichel2017largelyself}. The protocol is $\epsilon_{\text{QKD}}$-secure if:
\begin{equation}
\epsilon_{\text{QKD}} \leq \epsilon_{\text{PA}} + \epsilon_{\text{PE}} + \epsilon_{\text{EC}} + \epsilon_{\text{A}}.
\end{equation}
Here, $\epsilon_{\text{PA}}$ and $\epsilon_{\text{PE}}$ are the failure probabilities associated with Privacy Amplification and Parameter Estimation, respectively; together, they guarantee the secrecy of the key. The term $\epsilon_{\text{EC}}$ represents the failure probability of the Error Correction step, ensuring the correctness of the secret key.
Finally, we address the authentication of the classical communication channel described by $\epsilon_{\text{A}}$, which is the failure probability of the authentication. 

To ensure strict information-theoretic security, we adopt a conservative `worst-case' scenario, wherein every classical message exchanged during the protocol must be authenticated to prevent Man-in-the-Middle attacks. Consequently, the total authentication cost is determined by the cumulative length of the messages exchanged across five stages of the protocol:
\begin{enumerate}[nosep]
    \item \textbf{Measurement stage :} The declaration of measurement bases (relevant for Scenario 2).
    \item \textbf{Parameter Estimation:} Exchange of bits to estimate the QBER.
    \item \textbf{Error Correction:} Transmission of syndrome information to correct the key.
    \item \textbf{Error Verification:} Exchange of seed and hash values to verify key correctness.
    \item \textbf{Privacy Amplification:} Exchange of the seed to perform random extraction.
\end{enumerate}

Each of these steps consumes a portion of the initial secret key $K_{\text{HEPUF}}$. The sum of these costs determines the minimum required length $L$ of the HEPUF Authentication subroutine output. 

To authenticate the classical channel, we implement the WC authentication scheme. The authentication scheme proceeds as follows. Alice and Bob authenticate their classical communications using a family of Toeplitz matrices. Specifically, the parties utilize a $2t$-bit pre-shared secure key to select a hash function from this family, constructing a $k$×$n$ Toeplitz matrix via a Linear Feedback Shift Register (LFSR). The sender computes an authentication tag by multiplying this matrix by the message vector. To ensure the $2t$-bit matrix key remains secure against extraction and reusable for subsequent messages, the resulting tag is masked via encryption with an additional $k$-bit secure one-time pad. The sender transmits the message and the encrypted tag; the receiver applies the identical procedure, confirming the message's legitimate origin if the computed tags match \cite{Fung_2010}. The failure probability per authentication attempt is given by $\epsilon_{\text{A}}$. Following standard security bounds for polynomial hashing, this is given by:
\begin{equation}
\epsilon_{\text{A}} \leq n2^{-t+1},
\end{equation}
where $n$ represents the message length and $t$ denotes the length of the authentication tag. From this inequality, we can analytically deduce the lower bound on the required tag length $t$ to satisfy the overall security parameter $\epsilon_{\text{QKD}}$ for each authenticated message. The overall reliability of the authentication is characterized by its correctness bound, $\epsilon_{\text{A}}$, and its security bound, $\epsilon_{\text{HEPUF}}$. Specifically, $\epsilon_{\text{HEPUF}}$ quantifies the maximum probability that an adversary successfully compromises the authentication key. Provided the key remains secure and the protocol is executed without failures, the classical channel is authenticated. More details are provided in End Matter~\ref{sec:QKD Authentication}.

\begin{figure}[h]
\centering
{\includegraphics[width = 0.5\textwidth]{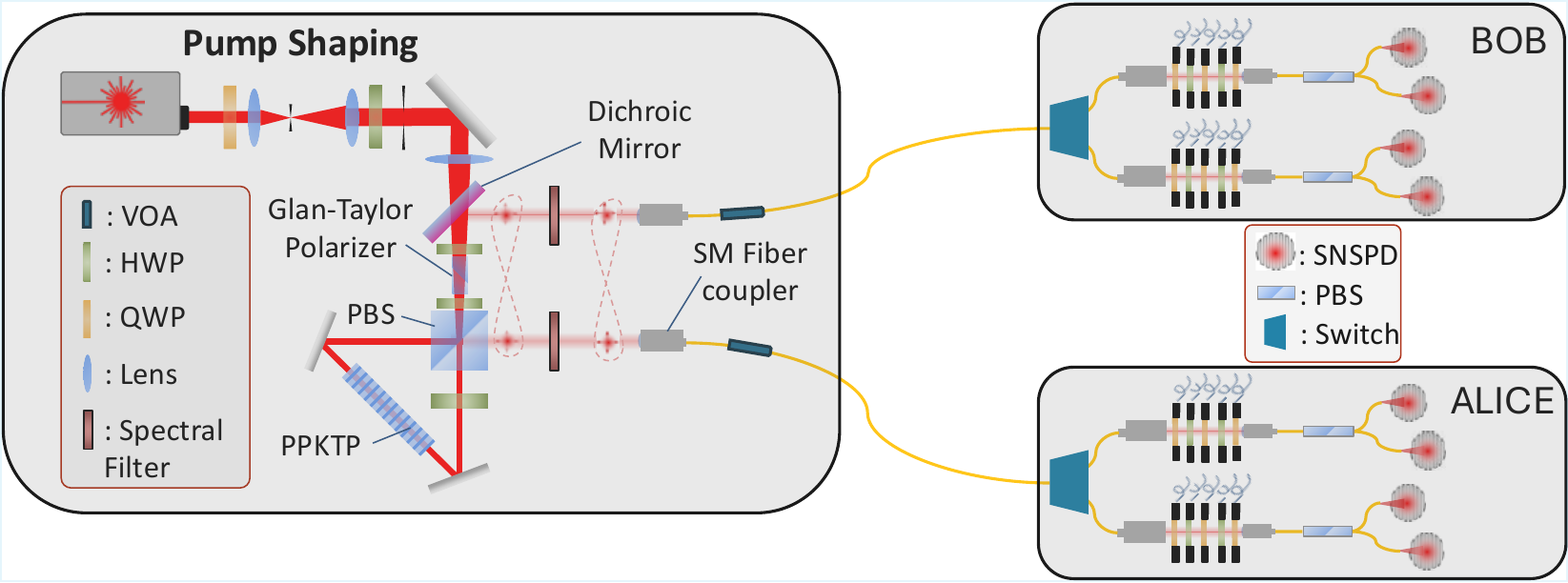}}
\caption{Experimental setup. A Ti:Sapphire laser (Coherent Mira-HP) with an average power of 3.4~W emits 2~ps pulses at a wavelength of 775~nm with a repetition rate of 76~MHz. The spatial mode of the laser is shaped into a Gaussian profile. The light entering the Sagnac interferometer is diagonally polarized to maximize the fidelity of output states with respect to the Bell state. Photon pairs are probabilistically generated via type-II Spontaneous Parametric Down Conversion in a ppKTP crystal (30mm-long, 46.2~$\mu$m poling period, provided by Raicol) and entangled in polarization in the Sagnac loop, resulting in the output state $(\ket{H}_s\ket{V}_i+e^{i\theta}\ket{V}_s\ket{H}_i)/\sqrt{2}$. After filtering the photons with a dichroic mirror and 1100~nm long pass filters, they are coupled to single mode fibers with a 12~mm focal lens. Three sets of QWP-HWP-QWP rotate the final state to the desired one. A set of HWP-QWP-FPBS is used to map each photon's polarization to the spatial degree of freedom. The fibers are connected to superconducting nanowire single-photon detectors (ID Quantique ID281) that, in turn, are linked to a time tagger that allows for counting and correlating detection events for the analysis. Two Variable Optical Attenuators (V15501 Thorlabs) are used in order to simulate fiber attenuation. Two Optical Switches (OSW22-1310E Thorlabs) are use to implement the protocol (see main text).}
\label{fig:setup}
\end{figure}

\textit{\textbf{Experimental Implementation—}} To implement the protocol, we employ a high-fidelity polarization-encoded Bell-state source based on Type-II spontaneous parametric down-conversion (SPDC) in a Sagnac interferometer configuration~\cite{dosSantosMartins:25}. The optical layout is shown in Fig.~\ref{fig:setup} and more details on the setup and on component characterization are provided in End Matter~\ref{sec:Experimental Setup}.

Following state generation, the photon pairs are distributed to the respective measurement apparatuses. We begin the experimental characterization by performing full quantum-state tomography to establish the baseline performance of the source. The reconstructed density matrix yields a Bell-state fidelity of $\mathcal{F} = 99.17 \pm 0.14$\,\%. This high fidelity corresponds to a low intrinsic QBER. While the raw source generation rate is approximately $300$\,kHz, the effective protocol speed in this proof-of-principle demonstration is constrained by active modulation components. In the HEPUF Authentication subroutine, the state choice is performed via a mechanically controlled quarter-half-quarter wave-plate sequence, limiting the HEPUF subroutine rate to $1$\,Hz. Similarly, the QKD subroutine utilizes an optical switch with a switching time of $0.5$\,ms, which would result in a maximum rate of $2$\,kHz. However, our characterization shows a total effective rate of $13$\,Hz for the QKD subroutine, due to connection delays introduced by the switch driver. We emphasize that these are technical limitations of the hardware used in our setup, not fundamental bounds of the protocol; utilizing GHz-bandwidth electro-optic modulators would readily bridge this gap.

We first evaluate the protocol in Scenario 1, demonstrating the versatility of the HEPUF subroutine. In this configuration, the HEPUF output determines the measurement bases making the subroutine adaptable to any assumptions. For the authentication phase, we emulate the Challenge-Response pair generation using a Permutation PUF model~\cite{Wisiol2019} to construct the Verifier's database. To satisfy the security criterion that we fix as $\epsilon_{\text{HEPUF}} < 2.5*10^{-11}$, our analysis dictates a sample size of $m = 44$ error-free states per authentication round, assuming at most $\delta \leq 0.1$. We execute the protocol over three different single channel attenuation levels: $15$\,dB, $20$\,dB, and $25$\,dB, (corresponding to a total attenuation of 30, 40, and 50 dB) performing $290$, $351$, and $351$ rounds, respectively and keeping  $\epsilon_{\text{QKD}} < 2.5*10^{-11}$. Note that the number of rounds is lower for the first attenuation level due to a technical issue during the measurement.

From the generated raw data, we can evaluate the experimental $\delta$. Here we will take the worst bias from each attenuation case, namely $\delta = 0.0016$. For error correction, we employ Low-Density Parity-Check (LDPC) codes with an estimated reconciliation efficiency of $f_{\text{EC}} = 1.06$. By applying the Leftover Hash Lemma and accounting for the finite-size effects and PUF bias, we derive the lower bound on the extractable secret initial key length $L$. The experimental results and key parameters are summarized in Table~\ref{tab:results_scenario_1}.
\begin{table}[h!]
\centering

\begin{tabular}{c c c c c}
\hline \hline
\textbf{Attenuation} & \textbf{$K_{\text{QKD}}$} & \textbf{$K_{\text{AUT}}$} & \textbf{QBER (\%)} & \textbf{SKR (bps)} \\
\hline
30 dB & 25094 & 237 & 0.56 & 0.381 \\
40 dB & 30000 & 237 & 0.52 & 0.310  \\
50 dB & 30000 & 237 & 0.55 & 0.270  \\
\hline \hline
\end{tabular}
\caption{Experimental results for Scenario 1 under a varying channel attenuation. The protocol parameters are set to ensure $\epsilon_{\text{HEPUF}} < 2.5*10^{-11}$, $\epsilon_{\text{QKD}} < 2.5*10^{-11}$ and $\epsilon_{\text{stat}} < 2.5*10^{-11}$ with a sample size of $m = 44$. $N$ denotes the number of rounds performed.}
\label{tab:results_scenario_1}
\end{table}

\begin{figure}[h]
    \centering
    \subfigure{\includegraphics[scale=0.47]{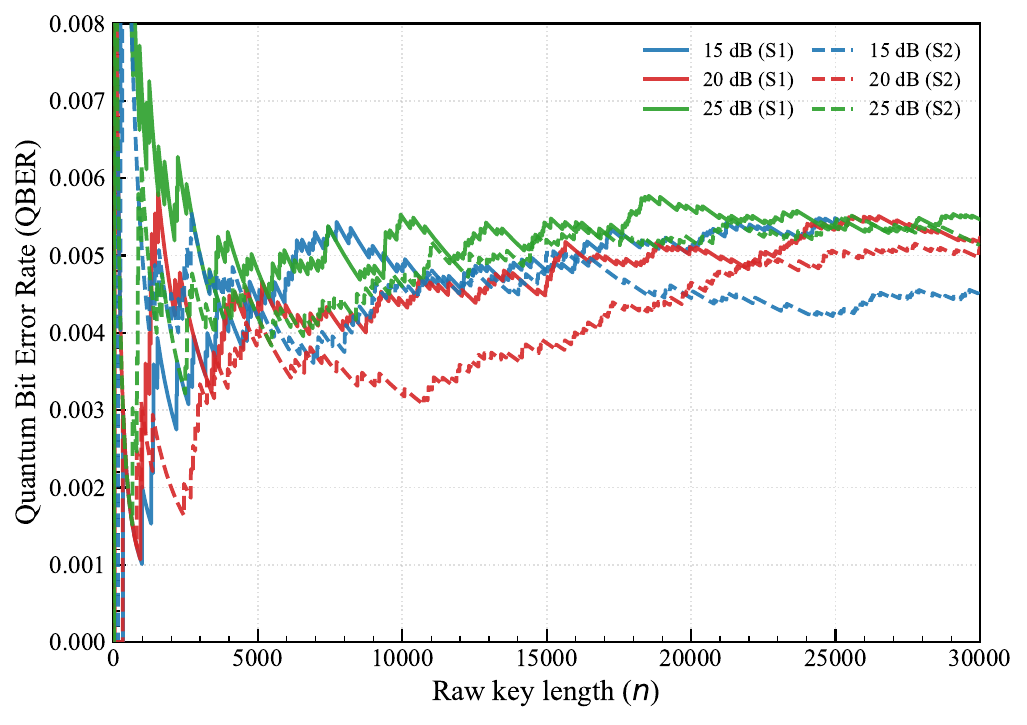}}
    \subfigure{\includegraphics[scale=0.4]{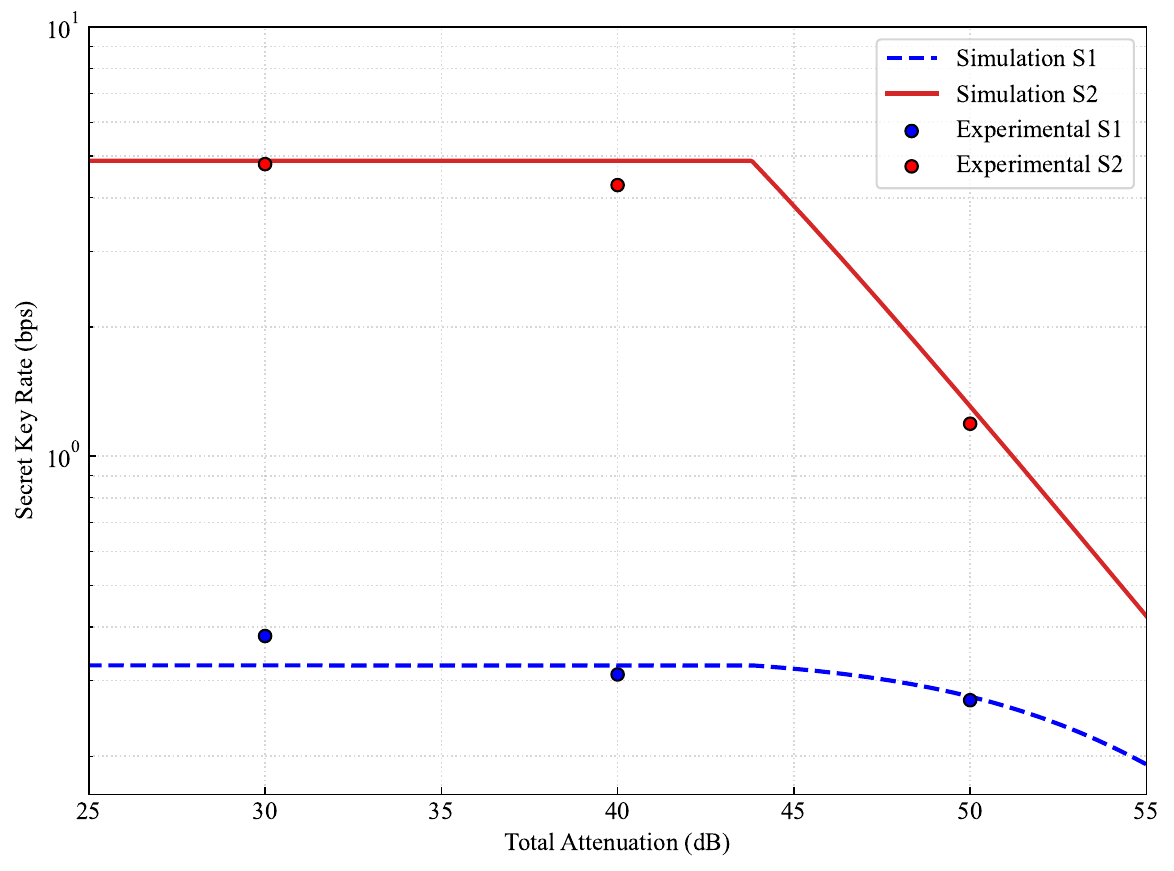}}
    \caption{Experimental performance of the HEPUF-authenticated QKD protocol. \textbf{Upper}: Evolution of QBER as a function of the raw-key length for attenuation levels of 30, 40, and 50 dB. Solid lines represent Scenario 1, while dashed lines represent Scenario 2. \textbf{Bottom}: SKR in bits per second versus channel attenuation. Blue markers denote Scenario 1 and red markers denote Scenario 2. Solid line represent simulation using experimental values.}
    \label{fig:QBER}
\end{figure}

As evidenced by Table~\ref{tab:results_scenario_1}, the system maintains a low QBER across all attenuation regimes, as also illustrated in Fig.~\ref{fig:QBER}. This stability is due to the high fidelity of the entangled photon source and confirms its suitability for long-distance (high-loss) entanglement-based QKD. As shown in Fig.~\ref{fig:QBER}, the QBER converges rapidly with the number of measured samples, further demonstrating the intrinsic stability of the source. While the absolute secret key rate is currently bounded by the mechanical actuation speed of the wave-plates -- as previously discussed -- the successful extraction of a secret key under these conditions validates the protocol's robustness. Collectively, these results constitute the first experimental demonstration of a QKD protocol authenticated via a Quantum Physical Unclonable Function. The rigorous derivation of the QKD security parameters and the finite-size key analysis are provided in detail in End Matter~\ref{sec:Finite key rate calculation}.

Next, we evaluate the protocol under Scenario 2, where by leveraging the inherent security of EB QKD, the parties are permitted to publicly disclose their measurement bases during the standard sifting phase. This relaxes the demand on the HEPUF subroutine: the initial secret key is only $K_{\text{AUT}}$ and is utilized exclusively for the WC authentication of the classical channel, rather than for the bit-expensive process of secret bases determination. As the key is smaller, the bias will be larger. We measure $\delta = 0.011$. 
We apply the same post-processing and estimate the initial secret-key length. The experimental results are summarized in Table~\ref{tab:results_scenario_2}. 
\begin{table}[h!]
\centering
\begin{tabular}{c c c c}
\hline \hline
\textbf{Attenuation} & \textbf{$K_{\text{AUT}}$} & \textbf{QBER (\%)} & \textbf{SKR (bps)} \\
\hline
30 dB & 270 & 0.45 & 4.79 \\
40 dB & 270 & 0.51 & 4.28 \\
50 dB & 270 & 0.51 & 1.19  \\
\hline \hline
\end{tabular}
\caption{Experimental results for Scenario 2 under a varying channel attenuation. The protocol parameters are set to ensure $\epsilon_{\text{HEPUF}} < 2.5*10^{-11}$, $\epsilon_{\text{QKD}} < 2.5*10^{-11}$ and $\epsilon_{\text{stats}} < 2.5*10^{-11}$  with a sample size of $m = 44$. $N$ denotes the number of rounds performed.}
\label{tab:results_scenario_2}
\end{table} Consistent with the first scenario, the system maintains a remarkably low QBER across all attenuation regimes, as shown in Fig.~\ref{fig:QBER}. Notably, the secret-key rate increases by approximately an order of magnitude compared to Scenario 1. This significant performance gain arises from the reduced consumption of the initial key, although the absolute throughput remains technically bounded by the total switching time.

\textit{\textbf{Discussion—}} In this Letter, we have reported on the first, to the best of our knowledge, experimental realization of Hybrid PUF (HEPUF) authentication integrated within an entangled Quantum Key Distribution system. We developed a unified framework tailored for two distinct operational regimes, demonstrating the versatility of the HEPUF as a secure quantum primitive. In the first scenario, we prioritize generality of integration with different cryptographic schemes. By utilizing the HEPUF output for both classical channel authentication and secret determination of measurement bases, we establish a protocol-agnostic solution capable of bootstrapping security without relying on computational assumptions. In the second scenario, we leverage the non-local correlations of entangled states to relax the bases secrecy requirement, thereby significantly enhancing the overall protocol efficiency. Experimentally, our setup demonstrates remarkable robustness. Thanks to a high-fidelity polarization-entangled photon pair source in a Sagnac configuration, we maintain an exceptionally low QBER even under high total channel attenuation ranging from $30$\,dB to $50$\,dB. These results not only validate the theoretical stability of the HEPUF architecture but also address one of the most critical, yet frequently overlooked, bottlenecks in practical QKD: the requirement for secure initial authentication. We believe that this work establishes the HEPUF as an important tool for quantum networks. By bridging the gap between hardware security and quantum communication protocols, our results pave the way for hybrid hardware-based modules into broader cryptographic applications, from device fingerprinting to secure multiparty computation.

\textit{\textbf{Acknowledgments—}} NLP and ED acknowledge financial support from the European Union’s Horizon Europe research and innovation programme under the project QSNP, grant no. 101114043, and of the PEPR integrated project QCommTestbed, ANR-22-
PET Q-0011, which is part of Plan France 2030. MD acknowledges the support of the UK Engineering and Physical Sciences Research Council, the Integrated Quantum Networks Hub, grant reference EP/Z533208/1, and the HSM-QCC QuantERA project, with the EPSRC grant reference EP/Z000564/1. AI acknowledges the support of the QuantERA QuantaGENOMICS project, under Grant Agreement No. 101017733.

\bibliography{references}

\clearpage
\section{HEPUF Architecture}
\label{sec:HEPUF Architecture}
In this section, we detail the internal structure of the HEPUF~\cite{2504.11552}. The device is composed of a classical $p$-randomness PUF~\cite{chakraborty_quantum_2021} inside a tamper-proof box, which also includes the encoding and measurement mechanism. The specific instance of the HEPUF used in this protocol generates entangled pairs from the set $\{\ket{\Phi^+},\ket{\Psi^-}\}$. The HEPUF operates in three distinct modes. The first mode (Mode 0) is exclusively accessible by the Verifier. By hardware assumption and construction of the HEPUF, this mode can only be used once (in a sequence of authentication rounds) in the setup phase to allow the generation of the classical Challenge-Response Database. After the setup phase, the HEPUF will be locked by setting the mode to 1 before being sent out to the Prover. The second mode (Mode 1) is then used in the authentication phase for the generation of the stream of entangled pairs with respect to the received challenge $x$. In this mode, the HEPUF internally uses classical PUF's outcome $y$, and the substring $y^2$ is encoded into the predefined bipartite states (per bit of $y^2$, if the bit is 0, the generated state is $\ket{\Phi^+}$, and if 1, it is $\ket{\Psi^-}$). The device retains the Prover's subsystem and transmits the corresponding subsystem to the Verifier. In the last mode (Mode 2), which is used in the verification phase, the HEPUF measures the Prover's subsystem using a measurement basis defined by the substring $y^1$ (the first half of the outcome of classical PUF). In the protocol, the Prover communicates the measurement outcome via the classical channel to enable the verification.


\section{Finite-key rate calculation}
\label{sec:Finite key rate calculation}
We used a finite-key rate analysis to evaluate the SKR. We define each security parameter: 
\begin{align*}
    &\epsilon_{\mathrm{PA}}(\nu) = \frac{1}{2} \sqrt{2^{-n\left(1-h_2(\Delta+\nu)\right)+r+z+\ell}} \\
    &\epsilon_{\mathrm{PE}}(\nu, \xi)^2 = 2\left(\exp \left(-\frac{2 b k \xi^2}{n+1}\right)+\exp \left(-2 \gamma\left[n^2 \nu^{\prime 2}-1\right]\right)\right) \\
    &\epsilon_{\text{EC}} = 10^{-(s+2)} \\
    &\epsilon_{\mathrm{QKD}} = 10^{-s},
\end{align*}
where $s$ is the general security parameter, $\nu$ and $\xi$ are deviation terms to optimize the secret-key rate and $z$ is the size of the syndrome used for error correction. $b$ is the sifted key length, $\Delta$ is the QBER. We solve the following algorithm: 
\[
\renewcommand{\arraystretch}{1.4} 
\begin{array}{ll}
\displaystyle \max _{\vec{x} \in \mathbb{R}^4} & \ell=\lfloor\alpha b\rfloor \\
\text{s.t.} & \epsilon_{\mathrm{EC}}+ \epsilon_{\mathrm{PE}}(\nu, \xi)+\epsilon_{\mathrm{PA}}(\nu) \leq \epsilon_{\mathrm{QKD}}, \\
& k=\lfloor\beta b\rfloor, \quad n=b-k, \\
& \gamma=\max \left\{\frac{1}{n+1}+\frac{1}{k+1}, \frac{1}{b_{\mathrm{err}}+1}+\frac{1}{b-b_{\mathrm{err}}+1}\right\}, \\
& b_{\mathrm{err}}=\lceil b(\delta+\xi)\rceil, \quad \nu^{\prime}=\nu-\xi, \quad n^2 \nu^{\prime 2}>1, \\
& \alpha \in[0,1], \quad \beta \in(0,1 / 2], \quad 0<\nu<\xi<1 / 2-\delta.
\end{array}
\]
Our objective is to find the maximum secret-key length $l$; thus, $\alpha$ here represents the secret-key rate. In addition, $k$ is the number of sifted bits allocated to parameter estimation, and $r = 1.06\,h_2(\delta)$ is the expected error correction leakage.

\section{QKD Authentication}
\label{sec:QKD Authentication}
For the authentication of classical channel of the QKD subroutine, we can calculate an upper bound on the minimun required initial key as follows: 
\begin{equation}
\begin{aligned}
\epsilon_A \leq\;&
n_{\text{BASIS}}
\left(2^{-t_{\text{BASIS}}+1} - 2^{-t_{\text{PA}}+1}\right) \\
&+ \frac{n_{\text{BASIS}}}{2}
\left(2^{-t_{\text{PE}}+1} - 2^{-t_{\text{EC}}+1}\right) \\
&+ n_{\text{SYN}}\,2^{-t_{\text{SYN}}+1},
\end{aligned}
\end{equation}
with $n_{\text{BASIS}}$, $t_{\text{BASIS}}$ being the number of measured bases and the size of the tag for those bases and $ n_{\text{SYN}}$, $t_{\text{SYN}}$ being the size of the syndrome and the size of the tag of the syndrome.

The total authentication cost is composed of three terms. The first term accounts for the overhead associated with the basis declaration, and the transmission of the authenticated seed for the privacy amplification process. The second term encompasses the bit exchange required for parameter estimation (QBER) and the seed used for error verification. Finally, the third term represents the cost to authenticate the syndrome information transmitted during the error correction phase. For each of these terms, we apply conservative upper bounds to ensure that that the protocol remains highly adaptable to varying network conditions. Notably, in Scenario~1, the authentication cost associated with basis declaration is zero as the measurement bases are derived directly from the secure HEPUF output, which reduces the initial secret-key consumption.

\section{Experimental Setup}
\label{sec:Experimental Setup}

\begin{figure}[!htbp]
\centering
{\includegraphics[width = 0.5\textwidth]{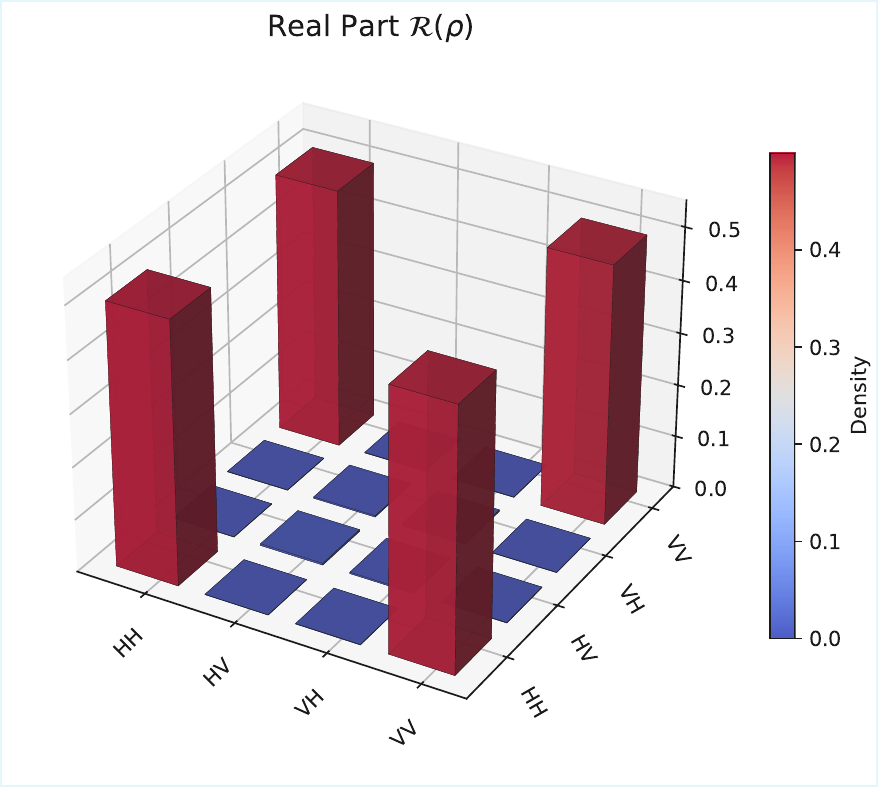}}
\caption{Experimental density matrix of the state produced with our source, estimated  via Quantum State Tomography, at $300$\,~kHz. The fidelity with respect to the Bell state is $99.17 \pm 0.14\,\%$ for an acquisition time of~$10$\,s per basis.}
\label{fig:density}
\end{figure}

The polarization-entangled state is produced via type-II spontaneous parametric down conversion (SPDC) occurring within a periodically-poled KTP (ppKTP) crystal. Polarization entanglement is achieved by pumping the crystal from two opposing directions and then interfering the two paths using a Polarizing Beam Splitter (PBS), realized with a Sagnac interferometer~\cite{PhysRevA.73.012316}. As a result, we obtain the Bell state $\ket{\Phi} = (\ket{HV} + e^{i\theta}\ket{VH})/\sqrt{2}$ (see Fig.~\ref{fig:density}), where $\theta$ is determined by the path difference between the two directions of propagation. With the aforementioned setup, we can generate a Bell state up to local unitaries resulting from the propagation of the state in single-mode fibers. To specifically produce any of the states of the set, we use an optimization method to determine the necessary local unitaries required to transform the state into the desired form. Using four sets of three Quarter-Wave Plates (QWPs), Half-Wave Plates (HWPs), and Quarter-Wave Plates (QWPs), we apply those unitaries to achieve the target state. More details can be found in~\cite{dosSantosMartins:25}. 

\end{document}